\documentclass[aip,superscriptaddress,showpacs,nofootinbib,
  floatfix,reprint]{revtex4-1}

\usepackage{amsmath}    
\usepackage{amssymb}
\usepackage{graphicx}   
\usepackage{verbatim}   
\usepackage{color}      
\usepackage{graphicx}
\usepackage{bm}
\usepackage{subfigure}  
\usepackage{hyperref}   

\usepackage{multirow}

\newcommand{\cch}[1]{\left[#1\right]}
\newcommand{\prt}[1]{\left(#1\right)}
\newcommand{\aver}[1]{\left\langle #1 \right\rangle}

\hypersetup{colorlinks=true, citecolor=blue,
  urlcolor=magenta,
  pdfcreator={pdflatex},
}

\begin{document}

\title{Order-disorder transition in a two-dimensional associating lattice gas}

\author{A. P. Furlan}
\email{apfurlan@fisica.ufmg.br}
\affiliation{Departamento de F\'isica, ICEx,
  Universidade Federal de Minas Gerais, C. P. 702, 30123-970 Belo
  Horizonte, Minas Gerais - Brazil}

\author{Tiago J. Oliveira}
\affiliation{Departamento de F\'isica, Universidade Federal de Vi\c cosa, 
36570-900, Vi\c cosa, Minas Gerais - Brazil}

\author{J\"urgen F. Stilck}
\affiliation{Instituto de F\'isica and National Institute of Science
  and Technology for Complex Systems, Universidade Federal Fluminense,
  Niter\'oi, Rio de Janeiro - Brazil}

\author{Ronald Dickman}
\affiliation{Departamento de F\'isica and National Institute 
of Science and Technology for Complex Systems, ICEx,
  Universidade Federal de Minas Gerais, C. P. 702, 30123-970 Belo
  Horizonte, Minas Gerais - Brazil}

\date{\today}

\begin{abstract}
    
  We study an associating lattice gas~(ALG) using Monte Carlo simulation on the triangular lattice and semi-analytical solutions on Husimi lattices. In this model, the molecules have an orientational degree of freedom and the interactions depend on the relative orientations of nearest-neighbor molecules, mimicking the formation of hydrogen bonds. We focus on the transition between the high-density liquid~(HDL) phase and the isotropic gas phase in the limit of full occupancy, corresponding to chemical potential $\mu \to \infty$, which has not yet been studied systematically. Simulation results show a continuous phase transition at 
  $\tau_c=k_BT_c/\gamma=0.4763(1)$ (where $-\gamma$ is the bond energy) between the low-temperature HDL phase, with a non-vanishing mean orientation of the molecules, and the high-temperature isotropic phase. Results for critical exponents and the Binder cumulant indicate that the transition belongs to the three-state Potts model universality class, even though the ALG Hamiltonian does not have the full permutation symmetry of the Potts model. In contrast with simulation, the Husimi lattice results furnish a discontinuous phase transition, characterized by a discontinuity of the nematic order parameter. The transition temperatures ($\tau_c=0.51403$ and $0.51207$ for trees built with triangles and hexagons, respectively) are slightly higher than the one found via simulation.  Since the Husimi
  lattice studies show that the ALG phase diagram features a discontinuous gas-HDL line for finite $\mu$, three possible scenarios arise. The first is that in the limit $\mu \to \infty$ the first-order line ends in a critical point; the second is a change in the nature of the transition at some finite chemical potential; the third is that the entire line is one of continuous phase transitions.  Results from other ALG models and the fact that mean-field approximations show a discontinuous phase transition for the three-state Potts model (known to possess a continuous transition) lends some weight to the third alternative.  
\end{abstract}

\maketitle

\section{Introduction}

It is no news that water exhibits quite unusual thermodynamic behavior, characterized by a set of anomalies, among which the anomaly in
density is the best known~\cite{Ch18}. In recent decades, many models were developed with the purpose of investigating the fundamental mechanisms which lead to the water anomalies. Among them, lattice models have attracted 
much attention due to their easy implementation and low computational 
cost. These models usually include soft-core potentials
to take care of excluded volume effects, and orientational 
interactions to represent hydrogen bonding between molecules. So far, these models are only able to reproduce some of the anomalies
qualitatively. In spite of this, they exhibit rich phase diagrams that provide an ideal environment for the study of phase transitions as well as the validation of new computational techniques. 

The first orientational lattice model for water was proposed by  Bell and Lavis~(BL)~\cite{Be70,Be702,La73}. It is defined on a triangular 
lattice, in which each site can be either vacant or occupied by a molecule.
The molecules possess three bonding directions with 120$^\circ$ between them, resulting in two orientational states per molecule. The model exhibits three phases: gas, low density liquid~(LDL) and high-density liquid~(HDL)~\cite{Fi09} at low, intermediate and high chemical potentials, respectively. While the gas-LDL transition is known to be discontinuous, characterized by a jump in density, the nature of the liquid-liquid transition is still controversial. For instance, mean-field approximations (Bethe lattice solutions~\cite{Ba08} and cluster-variation methods~\cite{Br02}) point to a discontinuous phase transition, whereas Monte Carlo simulations~\cite{Fi09,Se14} show a continuous transition. There is no consensus regarding the universality class of this transition. Fiore~{\it et al.}~\cite{Fi09} assert that it falls the Ising universality class, while 
\ifmmode \check{S}\else \v{S}\fi{}im\ifmmode \dot{e}\else \.{e}\fi{}nas
{\it et al.}~\cite{Se14} report three-state Potts critical exponents.

Following the ideas of Bell and Lavis, Henriques and Barbosa introduced a
two-dimensional~(2D) associating lattice gas~(ALG)~\cite{He05}. In this
model, also defined on a triangular lattice, each molecule has 
four bonding arms and two inert ones, the latter taking opposite directions on the lattice. Two of the bonding arms are 
proton donors in hydrogen bonds, while two are receptors, leading to eighteen states per molecule. The model~\cite{He05} also exhibits gas, LDL and HDL phases, but in contrast to simulation results for the BL model, the LDL and HDL phases are separated by a discontinuous transition line that ends at a bicritical point. There is also a gas-HDL line of continuous transitions, which starts at the bicritical point and seems to extend to large values of the chemical potential. The LDL and gas phases are separated by a continuous and a discontinuous transition line, which connect at a tricritical point. The former line meets the LDL-HDL and gas-HDL lines at the bicritical point~\cite{Sz09}.

Variants of the ALG model have been investigated, among them, 
three-dimensional (3D)~\cite{Gi07,Sz10,Fu16} and symmetric versions~\cite{Ba07,Fu15}. The 3D model~\cite{Gi07,Sz10} is defined on a body-centered cubic lattice in which each molecule possesses four bonding, and four inert arms. Despite the differences in geometry and number of orientational states in relation to the original ALG model~\cite{He05}, the three-phase (gas, LDL, HDL) behavior is preserved.  While a phase diagram featuring two tricritical points was suggested by Buzanno~{\it et al.}~\cite{Bu08} based on a cluster-variation approach, simulation results by Szortyka~{\it et al.}~\cite{Sz10} indicate that there is in fact a tricritical and a bicritical point, similar to the 2D case~\cite{Sz09}. In the 3D model, however, a gas-LDL coexistence line meets the continuous LDL-HDL and gas-HDL transition lines at the bicritical point~\cite{Sz10}.

The symmetric ALG model makes no distinction between donor and receptor bonding arms. This leads to a 
simplification, since the number of states per particle is substantially reduced. Balladares~{\it et al.}~\cite{Ba07} investigated this model on the triangular lattice and found only two discontinuous (gas-LDL and a LDL-HDL) transition lines in the phase diagram, each ending at a critical point. As an aside, let us remark that this very same scenario was reported in the former studies of the original 2D ALG model \cite{He05} and of the 3D version \cite{Gi07}. It turns out that they were incorrect, as demonstrated in more recent analyses~\cite{Sz09,Bu08,Sz10}, as discussed above. In fact, the semi-analytical solution of the symmetric model~\cite{Ba07} on a Husimi lattice build with hexagons (which is a mean-field approximation for the triangular lattice) unveils a phase diagram with three coexistence lines (gas-LDL, gas-HDL and LDL-HDL) meeting at a triple point~\cite{Ol11}. Moreover, more recent simulations of this model have provided evidence that the critical points reported in \cite{Ba07} are actually tricritical points~\cite{Fu15}, so that the thermodynamic behavior of this model is closer to the other versions. In contrast with these cases and with the mean-field results, however, a gas-HDL transition has not yet been observed in simulations of the 2D symmetric ALG, so that the existence of a continuous order-disorder transition and its universality class remains unclear for this model. In fact, studies of the critical exponents at the continuous phase transitions of all versions of the ALG model are still missing.  

Motivated by this issue, we study the phase transition in the full lattice limit of the symmetric 2D ALG model~\cite{Ba07}, using MC simulations (employing both Wang-Landau~\cite{Wa01,Wa012} and Metropolis algorithms), as well as obtaining the thermodynamic properties of the model in the core of Husimi cacti~\cite{Hu50}. A fully occupied lattice corresponds to the limit of infinite 
chemical potential, for which only the HDL and an isotropic, disordered phase   
(corresponding to what is identified as the gas phase in previous studies of the AGL) are expected, since the LDL phase becomes metastable already for small values of the chemical potential~\cite{Ba07,Ol11,Fu15}. 
Indeed, we find a single transition between the isotropic and the ordered HDL phase, whose loci, nature and universality class will be addressed in detail in the following.

The remainder of this paper is organized as follows. In section~\ref{sec:model} we detail the model. In Sec.~\ref{sec:details} we explain the simulation and analytic methods used. In Secs.~\ref{sec:sim_resul} and \ref{sec:hus_resul} we report our simulation results and Husimi-cacti findings, respectively. Finally in Sec.~\ref{sec:conc}, we discuss our conclusions and perspectives for future work.


\section{Model}\label{sec:model}
We consider the associating lattice gas~(ALG) introduced in Ref.~\cite{He05} in its 
symmetric version~\cite{Ba07}. The model was proposed in the context of water-like 
anomalies; despite its simplicity, it captures some features of liquid water, such 
as density~\cite{He05,Ba07} and diffusion~\cite{Sz09} anomalies. The model is defined 
on a triangular lattice (coordination number $z=6$) in which each site can be either 
empty or occupied by a molecule. Each molecule has six arms, four of which are bonding 
arms, while the other two are non-bonding (``inert"). In the symmetric version~\cite{Ba07}, 
all the bonding arms interact in the same manner, there being no distinction between 
proton donors and receptors. The inert arms do not interact and assume diametrically 
opposed positions, giving rise to three orientational states, ${\bm \eta}_i$ with $i=1,2$ 
or $3$. In this work we adopt a different notation from that of Refs. \cite{He05} and
\cite{Ba07}. We denote the generators of the triangular lattice by $\hat{e}_1$, 
$\hat{e}_2$ and $\hat{e}_3$, with,
\begin{equation}
  \hat{e}_1 = \bm{i}, 
\end{equation}
\begin{equation}
  \hat{e}_2 = +\frac{1}{2} \bm{i} + \frac{\sqrt{3}}{2} \bm{j}, 
\end{equation}
and
\begin{equation}
  \hat{e}_3 = -\frac{1}{2} \bm{i} + \frac{\sqrt{3}}{2} \bm{j}.
\end{equation}
\vspace{.5em}

\begin{figure*}[!htb]
  \includegraphics[scale=1.5]{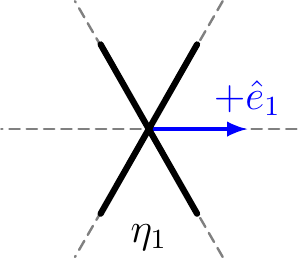} \hspace{1cm}
  \includegraphics[scale=1.5]{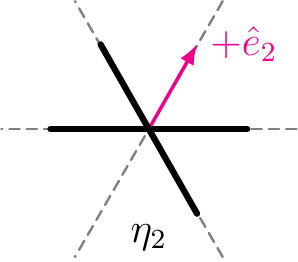} \hspace{1cm}
  \includegraphics[scale=1.5]{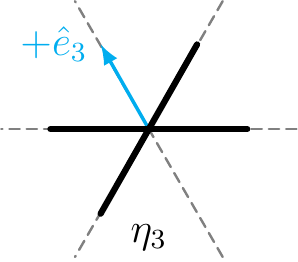}
  \caption{(Color online) Definition of orientational states ${\bm \eta}_1 \equiv \hat{e}_1$,
    ${\bm \eta}_2 \equiv \hat{e}_2$ and ${\bm \eta}_3 \equiv \hat{e}_3$
    respectively. The thick black lines represent bonding arms,
    dashed gray lines represent directions on the triangular lattice and
  the arrows indicate the generators of triangular lattice.}
  \label{fig:particle}
\end{figure*}

\noindent We use the same set of vectors to label the orientational states. Consider, 
for example, state 1, with bonding arms along $\pm \hat{e}_2$ and $\pm \hat{e}_3$. As
illustrated in Fig.~\ref{fig:particle}, we associate the vector ${\bm \eta}_1=\hat{e}_1$ 
with state 1, ${\bm \eta}_2=\hat{e}_2$ with state 2 and ${\bm \eta}_3=\hat{e}_3$ with 
state 3.  (Thus ${\bm \eta}$ points along one of the {\it nonbonding} directions.)
\vspace{1em}

In the ALG, interactions are restricted to nearest-neighbor
(NN) pairs, so that the separations ${\bm r}_i-{\bm r}_j$ between interacting pairs again
fall in the set $\{\pm \hat{e}_j\}$. A particle at site $k$ in state ${\bm \eta}_i$
has no interaction with its neighbors at sites ${\bm r_k}\pm \hat{e}_i$ since it
has no bonding arms pointing toward these sites. On the other hand, for $i \neq j$ we
have $|\hat{e}_i\cdot\hat{e}_j|=1/2$. Thus the interaction between a pair of
particles at sites $i$ and $j$, separated by ${\bm r}$ (a unit vector in the set
$\{\pm \hat{e}_i, i=1,2,3\}$), can be written so:
\begin{equation}
  \label{eq:interac}
  u_{ij}=-\gamma\prt{\frac{4}{3}}^2\cch{1 - \prt{{\bm \eta}_i \cdot {\bm r}}^2}
  \cch{1 - \prt{{\bm \eta}_j \cdot {\bm r}}^2},
\end{equation}
where $-\gamma$ denotes the energy associated with a NN particle pair with bonding arms
pointed toward each other. Using this, the energy of the ALG model may be written:
\begin{equation}
  \label{eq:hamil}
  {\cal H}\prt{{\bm \eta},{\bm r}} = \sum_{\aver{i,j}} \sigma_i \sigma_j [\varepsilon +  u_{ij}],
\end{equation}
where the summation $\aver{i,j}$ runs over pairs of nearest neighbors, the 
$\sigma_i=0,1$ are occupation variables, and $\varepsilon>0$ represents an
orientation-independent repulsive NN interaction.
In the present work, we consider full occupancy ($\sigma_i = 1, \forall i$), so that the parameter $\varepsilon$ is meaningless and $\gamma$ is the only remaining parameter in the Hamiltonian. Therefore, from here on we measure energy in units of $\gamma$, and define a dimensionless temperature $\tau=k_BT/\gamma$. We remark that when comparing our results with those in the literature for the full model \cite{Ba07,Ol11}, one should keep in mind that in these papers the particular case $\gamma/\varepsilon=2$ was considered, and the parameter $\varepsilon$ was used to define the reduced temperature and chemical potential. Thus, in these studies the reduced variables are twice those used here.

At full occupancy, each particle interacts with four of its six neighbors if all particles have the same orientational state (see Fig.~\ref{fig:particle}). Any configuration such that one or more pairs of nearest neighbors have distinct states must have a higher energy. Thus the ground state is threefold degenerate, with energy per particle (in units of $\gamma$), $e 
= -2$. By contrast, the mean energy per particle in a configuration in which each orientation 
is chosen at random, independently, is $e_{random} = - 4/3$.  Since the gain in entropy per 
site is $\Delta s/k_B = \ln 3$, a crude estimate for the critical temperature is 
$\tau_c \approx 2/(3 \ln 3) \simeq 0.6$. 

\begin{figure}[!htb]
  \includegraphics[scale=1.]{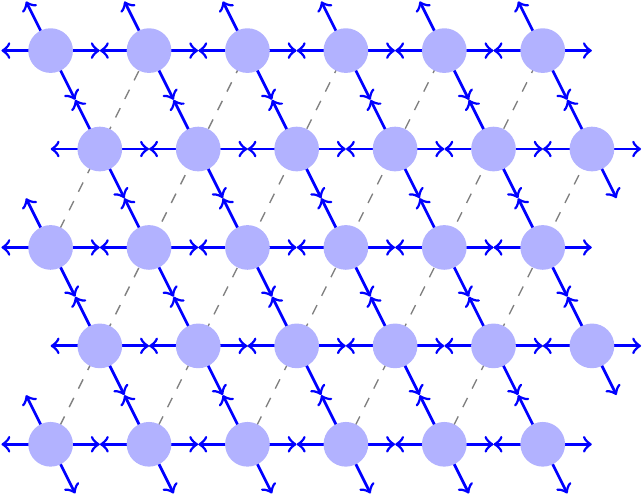}
  \caption{A ground-state configuration of the fully occupied lattice; all molecules are in the same state.}
  \label{fig:confs}
\end{figure}

The model studied here corresponds to the limit $\mu \to \infty$ of the full model, and at low temperatures it will be in the HDL phase.
Varying the temperature, we expect to observe a transition from the low-temperature ordered phase, with a majority of particles in one of the orientational states, to a high-temperature disordered one,
with equal populations among the three states. This isotropic phase corresponds to the gas phase in the general model. As noted in the Introduction, continuous gas-HDL phase transitions have already been observed in simulations of the nonsymmetric two-dimensional~\cite{Sz09} and three-dimensional~\cite{Bu08,Sz10,Fu16} ALG models, while a discontinuous transition was found in the mean-field approach for the symmetric model considered here~\cite{Ol11}. In the 2D case~\cite{Sz09,Ol11}, the transition is observed for reduced chemical potentials $\bar{\mu} \equiv \mu/\gamma \gtrsim 1$.
Szortika~{\it et al.}~\cite{Sz09} proposed an order parameter

\begin{equation}
  \label{eq:op2}
  \theta=\dfrac{3}{2}\prt{\frac{\max\prt{n_1,n_2,n_3}}{N}-\dfrac{1}{3}}
\end{equation}

\noindent where $n_i$ is the number of particles in state $i$ and $N$ is the total number of particles. Evidently, $\theta = 1$ when all
particles are in the same state, and
$\theta = 0$ for equally populated states.

In view of the orientation-dependent interaction and the orientationally ordered ground state, we can interpret the transition as one between nematic and isotropic phases.
Analysis of such a transition \cite{Plischke} suggests an alternative definition of the order parameter,

\begin{equation}
Q = \frac{1}{N} \sqrt{ n_1^2 + n_2^2 + n_3^2 - n_1 n_2 - n_1 n_3 - n_2 n_3} .
\label{eq:defQ}
\end{equation}

Of central interest is the nature of the order-disorder transition in the model at full occupancy.  (In principle, we would expect this also to describe the transition at large, but finite $\mu$.)  Since the ground state is threefold degenerate, it is tempting to suppose that the transition falls in the three-state Potts class.  While this may in fact be the case, we note that the energy, Eq.~(\ref{eq:hamil}), does not have the symmetry of the Potts model, that is, it is not invariant under permutations among the states.  (The ground state is permutation invariant, trivially, but due to the orientational dependence of the interactions, an arbitrary configuration is not.) 

In light of the above considerations, we regard it as an open question whether the model belongs to the 3-state Potts universality class.  The simulation results reported below provide some insight on this issue.

\section{Methods}\label{sec:details}

\subsection{Simulation details}
We performed extensive Monte Carlo (MC) simulations of the symmetric ALG model~\cite{Ba07} using both Wang-Landau (WL)
simulations~\cite{Wa01,Wa012} for triangular lattices with lateral sizes $24\leq L \leq 72$, and 
Metropolis simulations~\cite{Me53} for larger systems. The simulations are performed
at full occupancy ($N=L^2$).
The WL algorithm is an entropic sampling method, designed to estimate
$\Omega\prt{E}$, the number of configurations with energy $E$.
Starting from $\Omega\prt{E} = 1, \forall E$, the estimates for $\Omega$ are gradually
refined in a series of iterations.  Each iteration generates a sequence of
configurations as described below. An energy histogram
$H\cch{E}$, records the number of configurations having energy $E$.
These simulations use a random initial configuration
$\Gamma=\{{\bm \eta}_i\}$ with energy $E\prt{\Gamma}$. To generate a candidate for the
next configuration in the sequence ($\Gamma_{new}$), a site $i$
is chosen at random and its state is altered (${\bm \eta}_i \to {\bm \eta}^\prime_i$), generating configuration $\Gamma^\prime$.  This
is accepted as $\Gamma_{new}$ with probability

\begin{equation}\label{eq:probWL}
  p\cch{\Gamma \rightarrow \Gamma^\prime}=
  \min\cch{1,\dfrac{\Omega\cch{E\prt{\Gamma^\prime}}}
    {\Omega\cch{E\prt{\Gamma}}}}.
\end{equation}

\noindent With complementary probability, the current configuration $\Gamma$ is 
taken as the new configuration. The histogram and logarithm of the number of 
states updated so:
$H\cch{E\prt{\Gamma_{\rm new}}}\rightarrow H\cch{E\prt{\Gamma_{\rm new}}}+1$ and 
$\ln \Omega\cch{E\prt{\Gamma_{new}}} \rightarrow
\ln \Omega\cch{E\prt{\Gamma_{new}}} +\ln f$. 
The parameter $f$, known as the modification factor, is set to $e$, the base of 
natural logarithms, on the first iteration.  A sequence of configurations is 
generated by repeating this procedure, until the histogram is {\it flat}, i.e.,
there is no $E$ such that $H(E)$ is less than (or greater than) $X \%$ of the
mean value of the histogram over all energies.  Following the usual practice, we use $X=20\%$, that 
is, the $80\%$ flatness criterion.

Once the flatness criterion is satisfied, the current iteration ends and a new one 
is started with the histogram set to zero, but with the $\Omega\cch{E}$ carried
forward from the preceding iteration.  The new iteration proceeds as before except 
for a smaller
modification factor, taken as the square root of the previous value (i.e.,
$\ln f \to \ln f/2$). The usual procedure is to end the simulation once $\ln f<10^{-8}$ or equivalently after 27 iterations. 

As is evident from Eq.~(\ref{eq:probWL}), WL sampling is a kind of Metropolis importance sampling with target probability distribution $P(\Gamma) \propto 1/\Omega\cch{E(\Gamma)}$
in place of the usual Boltzmann distribution, $P(\Gamma) \propto \exp[-\beta E(\Gamma)]$.  Thus if iterations extended arbitrarily long and there were no sampling noise, the estimates for the $\Omega\prt{E}$ would converge to their true values.  For further details about the convergence (or lack thereof) of WL sampling, 
see~\cite{Be14} and references therein.

Recent studies~\cite{Ca14} show that it is not necessary to
use all 27 iterations of WL sampling. As the
modification factor is barely greater than unity, estimates
for the $\Omega\prt{E}$ are not significantly modified in the final few iterations.  We used
27 iterations for system sizes $L\leq 32$, 22 for $32 < L < 72$, 
and 20 for $L=72$.  

Using the estimates for the
$\Omega\prt{E}$,
the canonical average of a given thermodynamic observable ${\cal O}\prt{E}$ at
temperature $T$ can be computed via,

\begin{equation}
  \label{eq:avgcan}
  \aver{\cal O}\equiv\dfrac{1}{\cal Z}\sum_{E}\overline{{\cal O}\prt{E}}
    \Omega{\prt{E}}\exp\prt{-\beta E},
\end{equation}

\noindent where $\overline{{\cal O}\prt{E}}$ represents the microcanonical average,
and $\cal Z$ is the canonical partition function. 
Microcanonical averages are computed as simple averages of property ${\cal O}$ over all configurations
with energy $E$ generated in the final iteration of the WL procedure.
We compute the canonical average of the energy $E$ and its variance, and of order
parameters $\theta$ and $Q$ and their second through fourth moments.

Although WL sampling yields useful results for systems with $L \leq 90$, it
is not effective for larger systems.  (The time to achieve a flat histogram
becomes excessive.) 
We therefore use standard Metropolis sampling~\cite{Me53} for system sizes
$L=128$ and $L=256$. 
In these studies, we use $10^6$~MCS for equilibration
followed by $2 \times 10^6$~MCS to generate data. 
In both simulation methods, we perform an average over
60 independent realizations, starting from randomly generated initial configurations. 

\subsection{Mean-field approximations}

Generally, mean-field approximations of a
particular model correspond also to the solution of
this model on a complete graph, with properly
rescaled interactions~\cite{Ba82}. Since solutions of a
model on the core of a Cayley tree (a graph with no loops, in which each node has the same number of neighbors)
usually correspond
to the Bethe approximation of this
model on a regular lattice with the same 
coordination number as the tree, Baxter
suggested that these treelike lattices be
called {\it Bethe lattices}~\cite{Ba82}. Additional
correlations are taken into account considering {\it Husimi trees}, built with clusters (polygons or polyhedrons) rather than single sites, so that short closed paths are 
present. Analysis of a model on the core of a Husimi tree yields its
behavior on a {\it Husimi lattice}~\cite{Hu50}.

\begin{figure}
    \centering
    \includegraphics{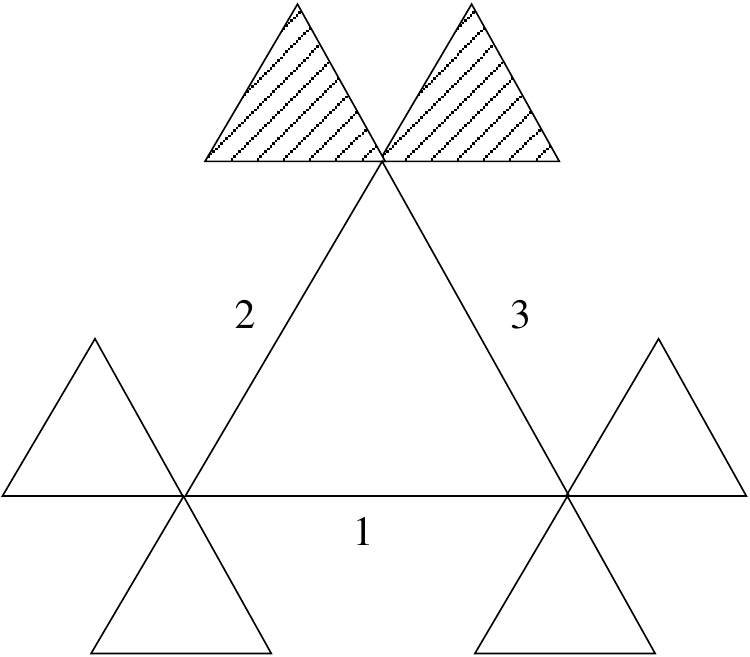}
    \caption{Six-coordinated Husimi tree built with triangles. Two generations of triangles are shown.  The
    numbers indicate the three directions of lattice edges, as defined in Fig. \ref{fig:confs}. The
    hatched triangles should be removed to define a sub-tree.}
    \label{fig:tree-tri}
\end{figure}

\begin{figure}
    \centering
    \includegraphics[scale=.55]{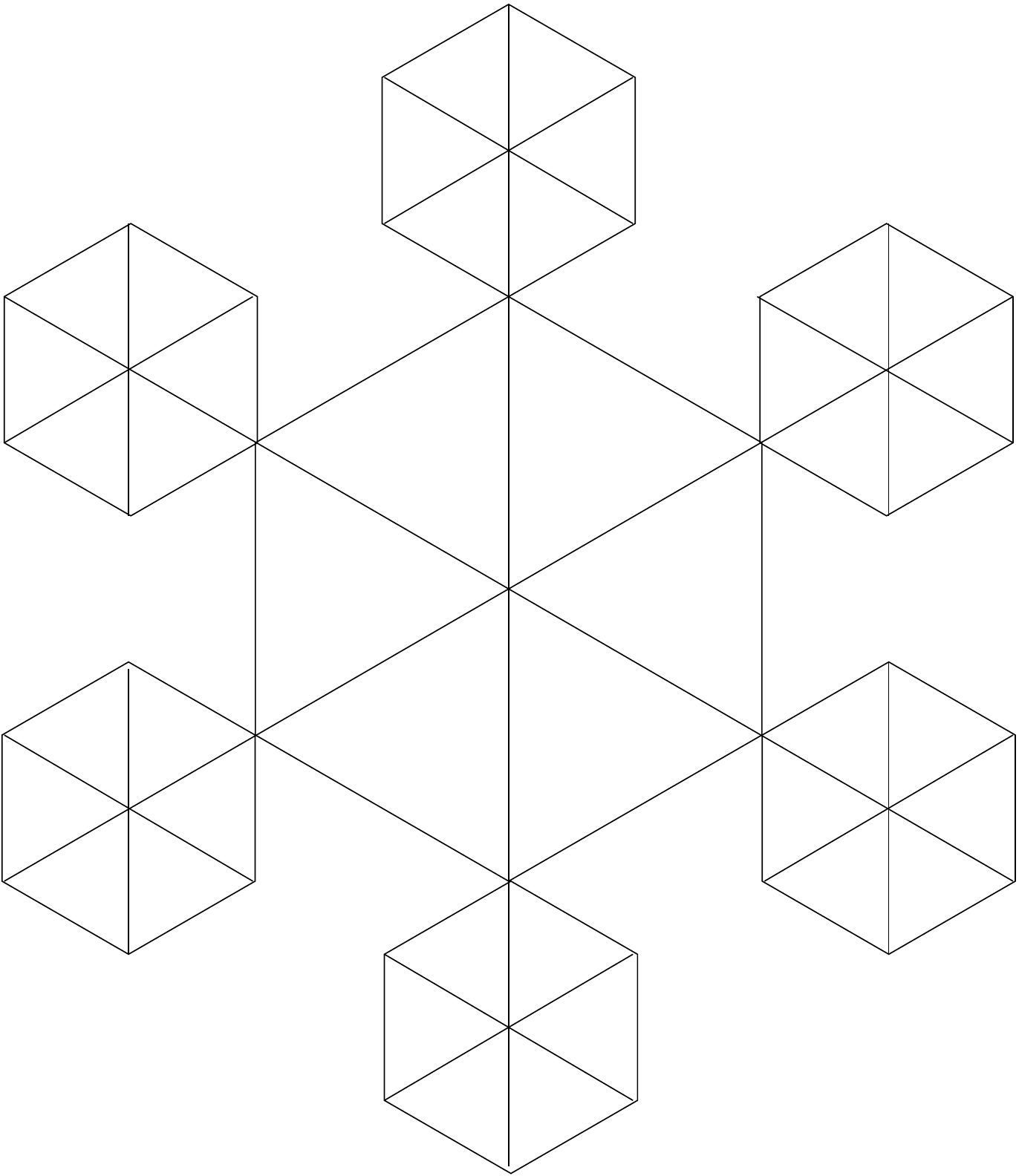}
    \caption{Husimi tree built with hexagons composed
    by six triangles. The central hexagon and the six hexagons of the
    first generation are shown.}
    \label{fig:tree-hex}
\end{figure}

We study the ALG model on such treelike lattices, in the limit in which each site 
is occupied by a molecule. 
Although we could begin with the six-coordinated Bethe lattice, one readily verifies 
that the ground state of the model on this 
lattice, without any closed paths, is highly degenerate. 
The thermodynamic behavior of the model on the Bethe lattice is
qualitatively distinct from that found on the triangular
lattice: there is no ordered phase. We therefore study
the model on a Husimi tree built with triangles,
in which three triangles meet at each site, as shown in Fig. \ref{fig:tree-tri}. We also study the model on
a more elaborate Husimi lattice built with
hexagons composed of six elementary triangles (see Fig. \ref{fig:tree-hex}). While the complete model, at finite chemical potential, was studied using similar techniques on this Husimi lattice~\cite{Ol11}, the limit ($1/\mu \rightarrow 0$) of interest here was not analyzed in that work.  As explained in Ref.~\cite{Ol11}, the use of hexagons in contrast to simple triangles as elementary clusters is imperative for capturing the symmetries of the ground states of the three phases of the full model (gas, LDL, and HDL) on the Husimi lattice; in the case of full occupancy, however, a tree of triangles (Fig. \ref{fig:tree-tri}) already captures the symmetry.

As usual, to solve a model on treelike lattices we define rooted sub-trees and their partial partition functions (ppf's) for the possible configurations of the site at the root. For the present
model, there are three possible orientations of the molecule
on the root site of a sub-tree, which may be represented by the 
variable $\eta$ assuming the values $1,2$ or $3$, depending on 
the orientation of the pair of non-bonding arms, as well as the
orientation of the root polygon. For example, in the tree built with
triangles, a sub-tree is
obtained if we remove the two hatched triangles in Fig. 
\ref{fig:tree-tri}, and the root site is opposite to the edge
with orientation 1. This defines the orientation of the
root triangle in this case as being equal to 1. A sub-tree with
an additional generation of polygons can then be built by connecting
two pairs of sub-trees to a new root triangle (or five sub-trees
to a new root hexagon on the second Husimi lattice). This 
operation leads to recursion relations for the ppf's. The 
configuration of the molecule at the root site is fixed and
one sums over the configurations of the molecules on the other
vertices of the root polygon. Upon iteration, the ppf's usually grow
without bound, and diverge in the thermodynamic limit. One thus
consider ratios of ppf's, which often reach  finite fixed-point
values. Thus, a thermodynamic phase is associated to one of the stable
fixed points of these ratios, and expected values of densities in the
central region of the tree may be obtained at the fixed points 
considering the operation of attaching three pairs of sub-trees to the
central triangle (or six sub-trees to the central hexagon in the second
lattice). Phase transitions are therefore related to distinct fixed
points competing in a region of the parameter space.
Finally, a free energy is needed to determine  the transition if coexistence of 
distinct phases occurs. To find the bulk free energy we resort to
an ansatz due to Gujrati~\cite{Gu95,Ol11}. The details of the 
calculations are presented in the appendix.

\section{Simulation results}\label{sec:sim_resul}

In this section we report simulation results for the model at full occupation.

The critical temperature $\tau_c$ is obtained via finite size scaling
(FSS)~\cite{Fi72} analysis of the susceptibilities $\chi_\theta$, 

\begin{equation}\label{eq:chi}
  \chi_\theta\prt{\tau}=\frac{L^2}{\tau}\cch{\aver{\theta^2} - \aver{\theta}^2},
\end{equation}
and $\chi_Q$, defined analogously, and of the specific heat $c$,

\begin{equation}\label{eq:cv}
  c\prt{\tau}=\frac{\beta^2}{L^2}\cch{\aver{E^2} - \aver{E}^2}.
\end{equation}

\noindent In Eqs.~(\ref{eq:chi}) and~(\ref{eq:cv}) the terms 
$\aver{\ldots}$ represent canonical averages, $L$ the linear size 
of the system and $\tau$ the temperature. (As mentioned in 
Sec.~\ref{sec:details}, our estimates for thermal averages such as 
$\theta$, $\chi$ and $c$ are averages over sixty independent realizations). 
The expected finite-size scaling forms for the order parameter,
susceptibility, and specific heat are,

\begin{equation}
  \theta \approx L^{-\beta_{\theta}/\nu}{\cal F}\prt{tL^{1/\nu}},
\end{equation}
and similarly for $Q$,
\begin{equation}
  \chi \approx L^{\gamma/\nu}{\cal X}\prt{tL^{1/\nu}},
\end{equation}
and
\begin{equation}
  c \approx c_{0}+L^{\alpha/\nu}{\cal C}\prt{tL^{1/\nu}},
\end{equation}

\noindent where $\beta$, $\gamma$, and $\alpha$ are critical exponents as
usually defined, and $t \equiv (\tau-\tau_c)/\tau_c$ is the reduced 
temperature. Additionally we determined Binder's fourth-order cumulant of 
the order parameters~\cite{Bi81},

\begin{equation}
  U_{4,\theta}(T) = 1 - \dfrac{\aver{\theta^4}}{3\aver{\theta^2}^2} .
\end{equation}
$U_{4,Q}$ is defined analogously.
At the critical point, $U_{4}$ tends to a universal value, 
characteristic  of the universality class, system shape and boundary conditions employed~\cite{Se07}.

\subsection{Determining \texorpdfstring{$\tau_c$}{Lg}}

\begin{figure}[!htb]
    \centering
    \includegraphics[scale=1.]{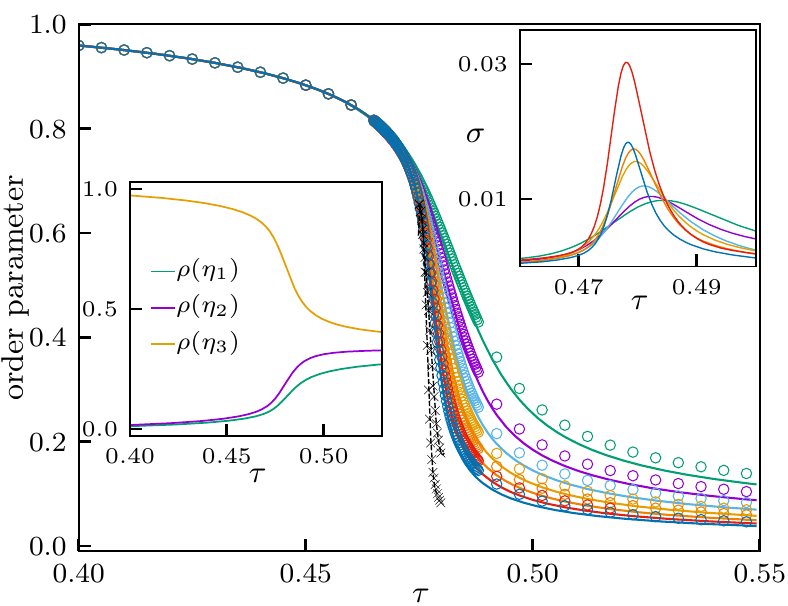}
    \caption{Order parameters $\theta$ and $Q$ versus temperature 
    for
    system sizes (upper to lower) $L=24,32,40,48,56,64,72,128$ and $256$.
    Solid lines and empty circles represent, respectively, 
    $\theta$ and $Q$ obtained via the WL algorithm. Black crosses
    represent results for $\theta$ obtained via Metropolis sampling 
    ($L=128$ and $256$); dashed lines represent polynomial fits to the 
    data. Lower inset: particle fractions $\rho$ in the majority (upper), 
    minority (lower) and intermediate states, for $L=32$. Upper inset: 
    relative uncertainties for $\theta$ in WL simulations; colors follow 
    the main plot.}
    \label{fig:HDLopxT}
\end{figure}

The results for the order-parameters $\theta$ and $Q$, plotted in
Fig.~\ref{fig:HDLopxT}, suggest that a continuous phase transition occurs at 
some temperature in the interval $0.45 < \tau < 0.5$.  The lower inset shows 
the fraction of particles $\rho_i$ in the majority, minority and intermediate
states, for system size $L=32$, illustrating a continuous variation between
ground state (all particles in the same state) and nearly equal populations.  
Relative uncertainties are plotted in the upper inset. As expected, they are
largest in the critical region. In all cases, the relative uncertainty in 
$\theta$ is less than 3\%; for system sizes other than $L=64$, it is
$<2$\%.

\begin{figure}[!htb]
    \centering
    \includegraphics[scale=1.0]{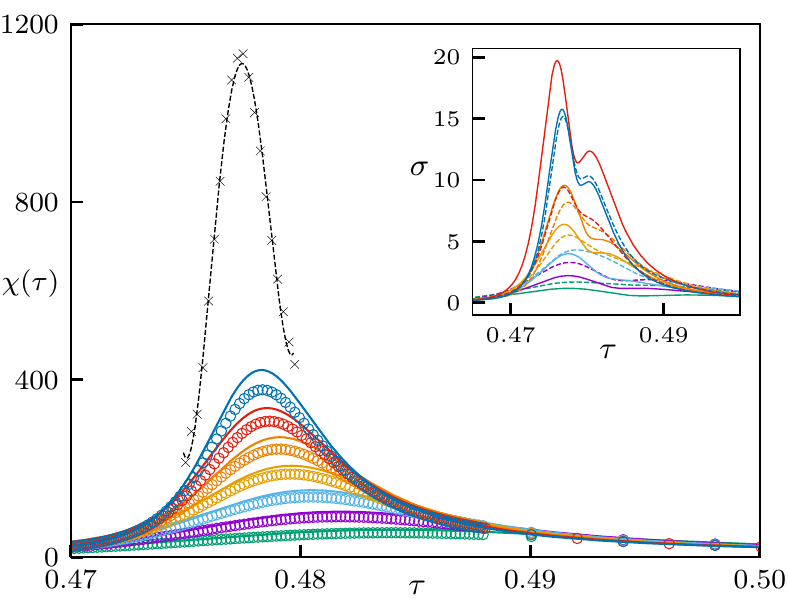}
    \caption{Susceptibilities versus temperature
        for system sizes $L=24,32,40,48,56,64,72$ and $128$. Solid
        lines and empty circles represent, respectively, $\chi_\theta$ and 
        $\chi_Q$ obtained via the WL algorithm. Black crosses   
        represent results for $\chi_{\theta}$ obtained via Metropolis simulations ($L=128$); 
        dashed lines represent a polynomial fit to the data for $L=128$. (For
        better visibility, the data for $L=256$ are not shown.)
        Inset: uncertainties in $\chi_{\theta}$ (solid lines)
        and $\chi_Q$ (dashed lines).} 
    \label{fig:HDLxixT} 
\end{figure}

\begin{figure}[!htb]
    \centering
     \includegraphics[scale=1.0]{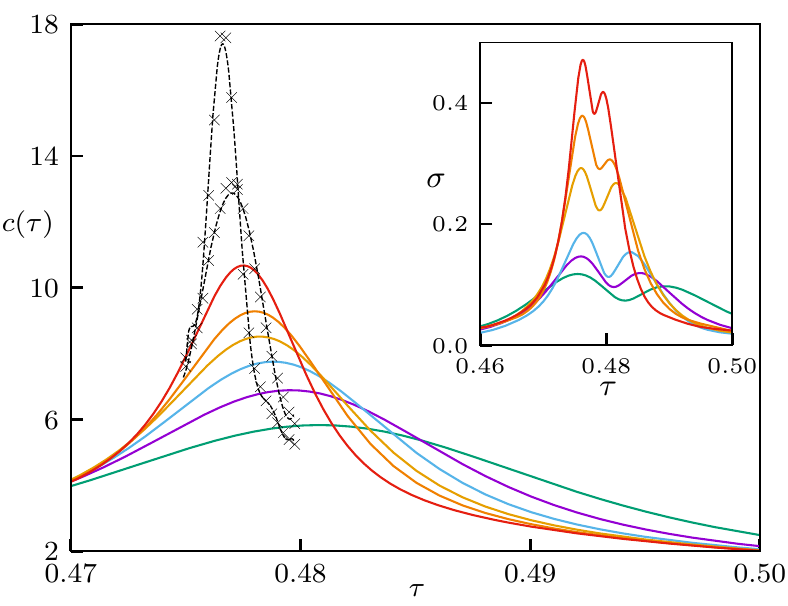}
    \caption{Specific heat versus temperature for system sizes
        $L=24,32,40,48,56,64,72,128$ and $256$. Solid lines
        represent results from WL simulations and black crosses 
        results obtained via Metropolis sampling for $L=128$ and 256. Dashed lines represent a 
        polynomial fit to the simulation data. Inset: 
        uncertainties in $c$.
        }
    \label{fig:HDLcvxT} 
\end{figure}

The susceptibilities $\chi_\theta$ and $\chi_Q$ are plotted in
Fig.~\ref{fig:HDLxixT}. Although these quantities exhibit similar behaviors, 
there are slight differences in the critical region; the differences are 
more evident for larger systems. The discrepancies between the susceptibilities 
are related to their structures. While $Q$~(see Eq.~\ref{eq:defQ}) takes into
account the densities of all three states, $\theta$ only involves the majority
density~(see Eq. \ref{eq:op2}). This may be why the  uncertainties~(inset
Fig.~\ref{fig:HDLxixT}) in $\chi_Q$ are approximately one-third those in
$\chi_\theta$. For the sizes shown in Fig.~\ref{fig:HDLxixT}, the maximum 
relative uncertainty $\chi_{\theta}$ in is about $6\%$.  
The specific heat $c$, shown in Fig.~\ref{fig:HDLcvxT},  
exhibits behavior consistent with that of the order parameters and 
susceptibilities. Relative uncertainties in $c$ are smaller than $4\%$ for 
$48 \leq L \leq 74$ and smaller 
than $2\%$ for $L<48$.

\begin{figure}[!htb]
    \centering
    \includegraphics[scale=1.]{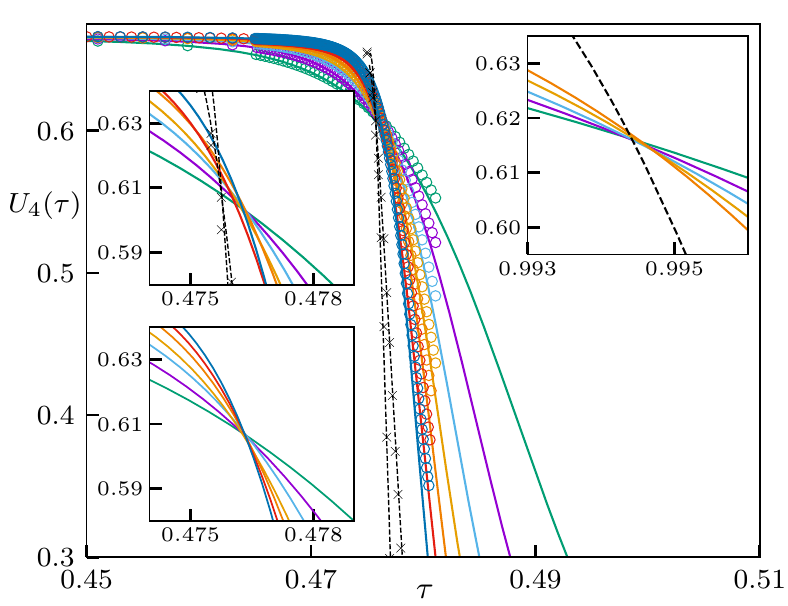}
    \caption{Binder cumulants $U_{4,\theta}$ (solid lines) and  
        $U_{4,Q}$ (circles) versus temperature for system sizes
        $L=24,32,40,48,56,64,72,128$ and $256$. Crosses: results for $U_{4, \theta}$
        using Metropolis sampling for $L=128$ and 256; dashed lines: sixth-order 
        polynomial fits to the simulation data. Left inset: detail 
        of the crossing region of $U_{4,\theta}$;  
        upper-right inset: detail 
        of the crossing region of $U_{4,Q}$. 
        Lower-right inset: cumulant of the three-state Potts model on the square lattice for 
        sizes 
        $L=32,40,48,56,64$ and $128$;
        the dashed line represents $L=128$.} 
    \label{fig:HDLu4xT}
\end{figure}

The Binder cumulants of the order parameters are shown
in Fig.~\ref{fig:HDLu4xT}. 
The crossings of the Binder cumulants for $\theta$ and $Q$ provide the
estimates $\tau^{(Q)}_c=0.476(4)$ and $\tau^{(\theta)}_c=0.476(5)$,
respectively. 

Using our results for $c$, $\chi_\theta$, $\chi_Q$, and the cumulants, we
estimate the critical temperature, $\tau_c$.  For $c$ and the $\chi$'s,
we define a size-dependent {\it pseudocritical temperature} as the 
temperature associated with the maximum value. Pseudocritical temperatues 
for the cumulants are identifying as the crossing temperatures 
of $U_{4}$ between: 1) the smallest system size studied, $L=24$, and the
others~($L=32,\ldots,L=256$) and 2) the crossing temperatures between a 
given system size $L$ and the next system size, for example, $L=24$ with
$L=32$, $L=32$ with $L=40$ and so on. The results from the adjacent sizes
crossings are represented as $U^\prime_4$.

\begin{figure}[]
    \includegraphics[scale=1.]{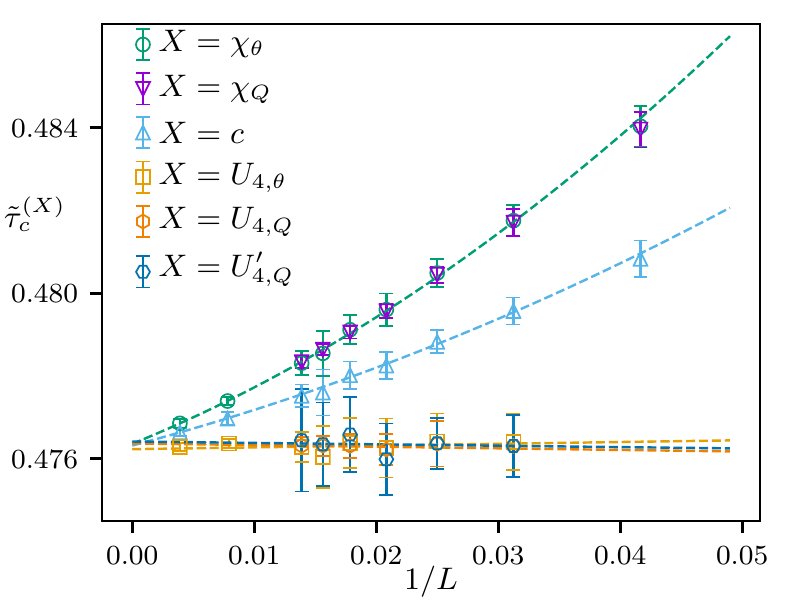}
    \caption{Pseudocritical temperatures associated with
    $\chi_\theta$,$\chi_Q$,$c$,$U_{4,\theta}$ and $U_{4,Q}$, 
    with symbols as indicated, versus $1/L$. Broken lines are 
    fits to the data (quadratic, except for the  cumulant data, 
    for which linear fits are very good).}
    \label{fig:HDLTc}
\end{figure}

The pseudocritical temperatures are plotted versus $1/L$ in
Fig.~\ref{fig:HDLTc}.  All six sets of pseudocritical temperatures 
appear to converge to similar values as $L \to \infty$. Of note is 
the relative insensitivity to system size of the pseudocritical
temperatures derived from cumulant crossings. The crossings between Binder 
cumulant of two adjacent system sizes for the order parameter $\theta$ 
suffer from large uncertainties for $L \geq 64$, affecting the estimate 
of the pseudocritical temperature. For this reason, we disregard this property 
in the calculation of the critical temperature. 

The resulting estimates for $\tau_c$ are listed in Table~\ref{tab:critvals}.
From the six estimates for the pseudocritical temperatures we derive the
global estimate $\tau_c = 0.4763(1)$. The global estimate of $\tau_c$ was
obtained through a weighted average of the properties with weights
$1/\sigma^2$, where $\sigma$ represents the uncertainty of each quantity.
\begin{table*}[htb]
\begin{tabular}{ccccccc}
    \hline\hline
      $c$    & $\chi_\theta$ & $\chi_Q$ & $U_{4,\theta}$ & $U_{4,Q}$
      & $U^\prime_{4,Q}$ & $\tau_c$ \\ \hline
      0.47632(1) &  0.47636(1) &  0.4761(1)   &  0.47621(2)  & 0.47637(8)
      & 0.4764(5) & 0.4763(1)  \\ \hline\hline
\end{tabular}
\caption{Estimates for the critical temperature. The first six columns 
(second line) list the estimates for $T_c$ obtained from analysis of 
the quantities listed on the first line, while the seventh column
contains the global estimate and its uncertainty.}
\label{tab:critvals}
\end{table*}

\subsection{Critical exponents}

Extrapolating the cumulant crossings to infinite size yields two
estimates for the critical cumulant: $U^\ast_{4,\theta}=0.604(2)$ 
and $U^\ast_{4,Q}=0.606(1)$. Since there can only be one value, 
a more conservative estimate is $U^*_4 = 0.605(2)$.  It is of interest
to compare this to the critical cumulant of the two-dimensional,
three-state Potts model. The best available estimate of the latter,
to our knowledge, is~\cite{To02} $U^*_{\rm Potts}=0.61$. 
(In Ref. \cite{To02}, no uncertainty estimate is provided; we may 
assume it is on the order of 0.01.) In efforts to improve on this
estimate we simulated the 3-state Potts model on $L \times L$ square lattices with
periodic boundaries. The results for $U_{4,{\rm Potts}}$, shown in
Fig.~\ref{fig:HDLu4xT} (lower-right insert), yield 
$U^\ast_4=0.6124(8)$ for the critical cumulant.  Although strictly
speaking incompatible with our estimate for the associating lattice
gas, the two estimates are rather similar, leaving open the
possibility of a common universality class for the two models.

\begin{figure}[!htb]
    \includegraphics[scale=1.]{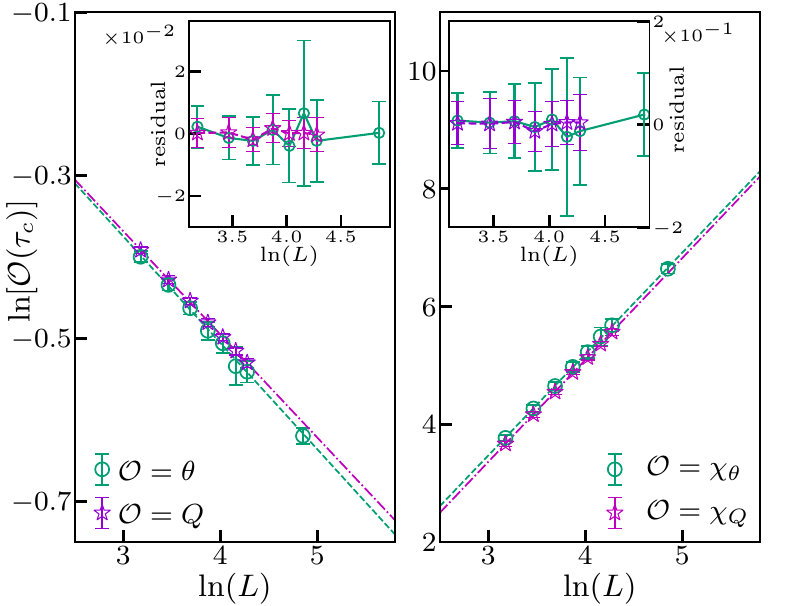}
    \caption{(Left) Order parameters 
    $\theta$($Q$) versus system size. The 
    circles(stars) represent simulation data; 
     dashed(dot-dashed) lines are least-squares
    linear fits of the data. Insets show the residuals 
    of the fits. Graphs on the right are analogous plots for the 
    associated susceptibilities.}
  \label{fig:expall}
\end{figure}

We estimate the critical exponent ratios $\beta/\nu$ and $\gamma/\nu$ 
using fits to the data for $\theta$, $Q$, and the
associated susceptibilities $versus$ system size (see Fig.~\ref{fig:expall}).

From the linear fits shown in Fig.~\ref{fig:expall} (left) and 
including the effect of the uncertainty in $\tau_c$ we obtain
$\beta_\theta/\nu=0.130(3)$ and $\beta_Q/\nu=0.126(1)$ 
The inset shows that the residuals are much smaller than the uncertainties, 
indicating that adding additional terms to the fitting function, in the 
form of corrections to scaling, would not yield better
estimates. The value of this ratio for the two-dimensional, three-state 
Potts model is~\cite{Wu82,Na13,Ca14}
$\beta_{\rm Potts}/\nu_{\rm Potts}=2/15=0.1333\ldots$. In the 
worst case ($\beta_Q$), the discrepancy is about $4\%$.
Linear fits to 
$\ln \chi_\theta$ and $\ln \chi_Q$ as functions of $\ln L$ 
(see Fig.~\ref{fig:expall}(right)) yield $\gamma_\theta/\nu=1.70(2)$ 
and $\gamma_Q/\nu=1.73(9)$. Our estimate for $\gamma_\theta$ differs 
from the value of the three-state Potts model, 
$\gamma_{\rm Potts}/\nu_{\rm Potts}=26/15=1.733\ldots$, by $1.7\%$. 
Analysis of the residuals of fits to the susceptibilities again 
indicates that adding further terms to the fitting function is 
not necessary. 

\begin{figure}[htb]
    \centering
    \includegraphics[scale=1.]{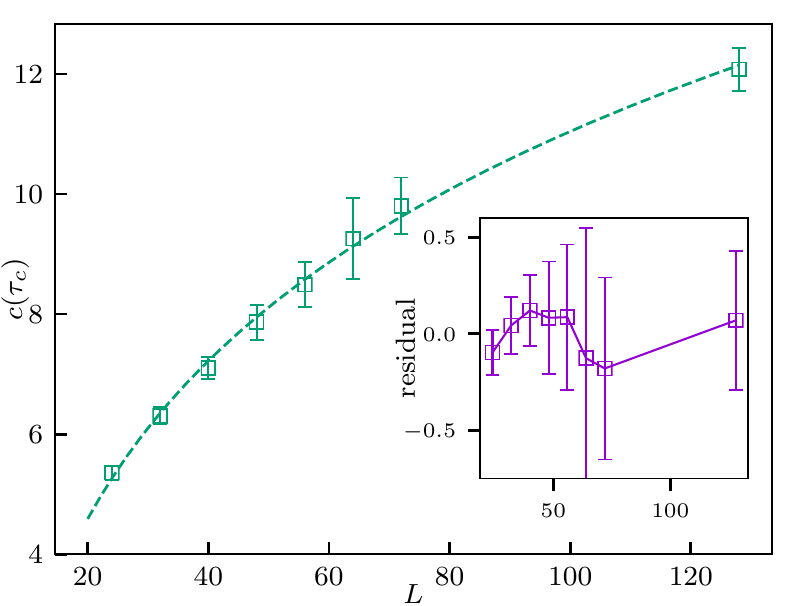}
    \caption{Specific heat versus system size at critical
    temperature. Squares: simulation data with their respective 
    uncertainties; dashed line: fit, $c=aL^{\alpha/\nu} + c_0$. 
    The inset shows the residuals, with error bars.}
    \label{fig:cvadjust}
\end{figure}

Different from the order parameter and susceptibility, a linear fit to 
$\ln c$ as a function of  $\ln L$ does not yield a good description of the 
data.  We therefore fit the data with $c=aL^{\alpha/\nu} + c_0$ with $a$, 
$\alpha/\nu$ and $c_0$ as adjustable parameters. This form is capable of 
fitting the simulation data, as can be seen in Fig.~\ref{fig:cvadjust}. 
The inset shows that the residuals are smaller than the uncertainties, 
and do not exhibit a systematic tendency. A least-squares procedure yields 
$a=2.4(6)$, $\alpha/\nu=0.38(3)$ and $c_0=-3\pm 1$.  The exponent ratio 
agrees to within uncertainty with that of the three-state Potts model, 
$\gamma/\nu_{\rm Potts}=0.4$.

\noindent Given the scaling relation $\alpha + 2\beta + \gamma = 2$, we have,

\begin{equation}
    \nu = \frac{2}{\alpha' + 2 \beta' + \gamma'},
    \label{eq:nus}
\end{equation}
where exponents with primes denote the corresponding exponents 
divided by $\nu$, determined in the finite-size scaling analysis 
discussed above.

Using the values for $\beta'$ and $\gamma'$ obtained using the order 
parameter $\theta$, equation~\ref{eq:nus} yields $\nu_\theta=0.85(5)$, 
while the values associated with order parameter $Q$ furnish 
$\nu_Q=0.84(1)$, for an average of $\nu = 0.85(1)$, close to the Potts 
model value, $\nu_{\rm Potts}=5/6=0.833\ldots$. 

\begin{figure}[htb]
    \centering
    \includegraphics[scale=.7]{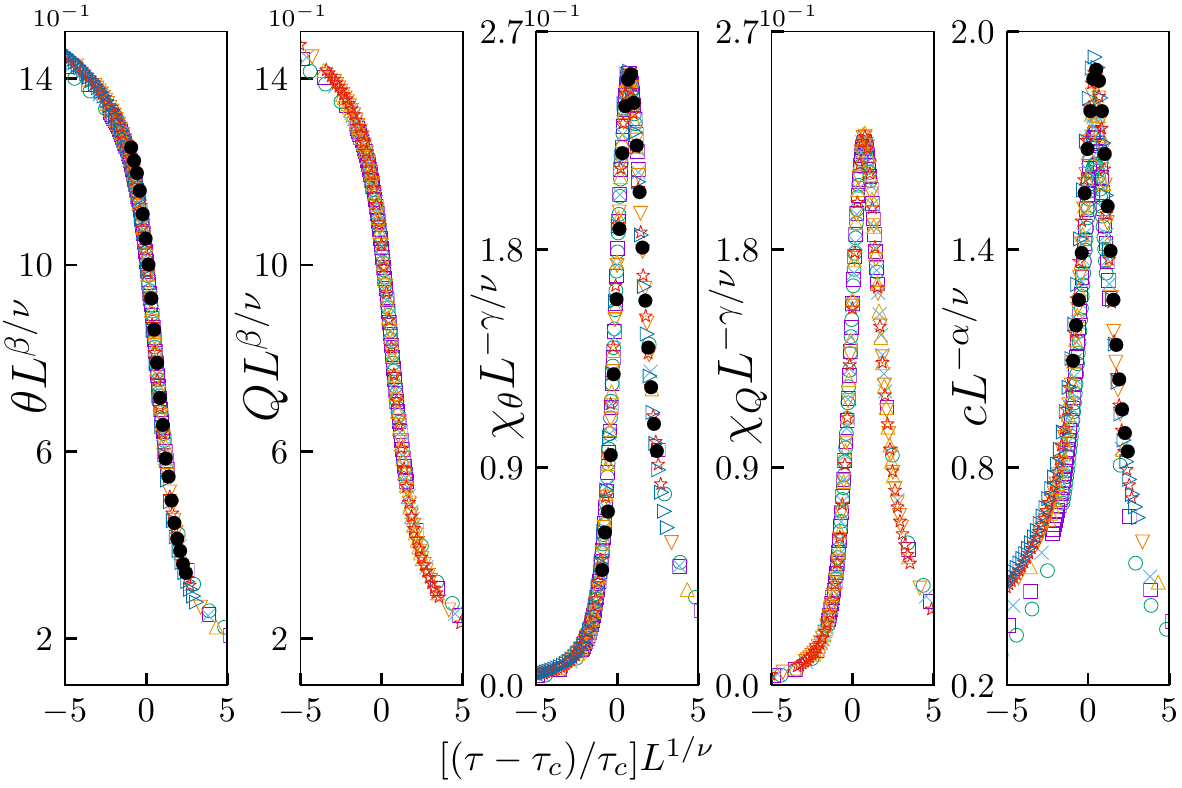}
    \caption{Data collapse for $\theta,Q,
    \chi_\theta,\chi_Q$ and $c$ respectively.
    The symbols $\bigcirc$,$\square$,$\times$,
    $\bigtriangleup$,$\bigtriangledown$,$\star$ and 
    $\rhd$ represent $L=24,32,40,48,56,64,72$ respectively.  
    Filled black circles ($\bullet$) represent $L=128$. } 
    \label{fig:collapse}
\end{figure}

As a further test regarding 
the universality class, we perform a data collapse using 
the Potts critical exponents, as shown in
Fig.~\ref{fig:collapse}.
The quantities $\theta,Q,\chi_\theta$ and
$\chi_Q$ exhibit a good collapse using these
exponents. We observe that the specific heat exhibits the poorest
collapse. 

The results for the exponents $\alpha, \beta$, $\gamma$ and
$\nu$, as well as for the Binder cumulant strongly suggest that the
phase transition studied here belongs to the three-state Potts
model universality class. Deviations of the critical exponents
from the Potts class values may be due to the flatness
criterion, limited sample size, and restricted system sizes.


\section{Results on Husimi lattices}\label{sec:hus_resul}

The thermodynamic behavior of the model on the core of both treelike lattices will be
presented here, after a brief description of the calculations which lead to it.
Some more details of these calculations may be found in the appendix.
As mentioned in Sec. \ref{sec:details}, recursion relations for the partial
partition functions on a rooted sub-tree are obtained, and 
ratios of these fpp's usually remain finite as the recursion relations are 
iterated. The stable (real and positive) fixed points attained iterating the recursion relations for the ratios
correspond to the thermodynamic limit. The behavior on 
the complete tree may then 
be found connecting sub-trees to a central polygon. Expected values of densities
at this central polygon may then be calculated, and may be seen as approximations
to the results on the triangular lattice. 

Two stable fixed points are found. At higher temperatures, only a disordered 
{\em isotropic} fixed point is stable, for which the densities of molecules in the
three possible orientations are equal. At low temperatures, three ordered 
{\em nematic} fixed points are stable, where the density of molecules in one of
the three orientations is larger than the ones in the other two orientations. The 
stability of these fixed points may be studied using the Jacobian of the 
recursion relations for the ratios. It is found that there is an interval of
temperatures where both kinds of fixed point are stable, signalling that the 
transition between the phases is discontinuous. Thus, to find the 
coexistence temperature the free energy of both phases should be equated. It should
be remembered that this free energy corresponds to the model on the {\em core} of
the tree; the free energy of the whole tree is dominated by the surface
\cite{Ba82,Gu95}. We find a coexistence temperature $\tau_c=0.51403$ for the 
triangle tree and $\tau_c=0.51207$ for the one with hexagons. 

\begin{figure}
    \centering
    \includegraphics[scale=1.]{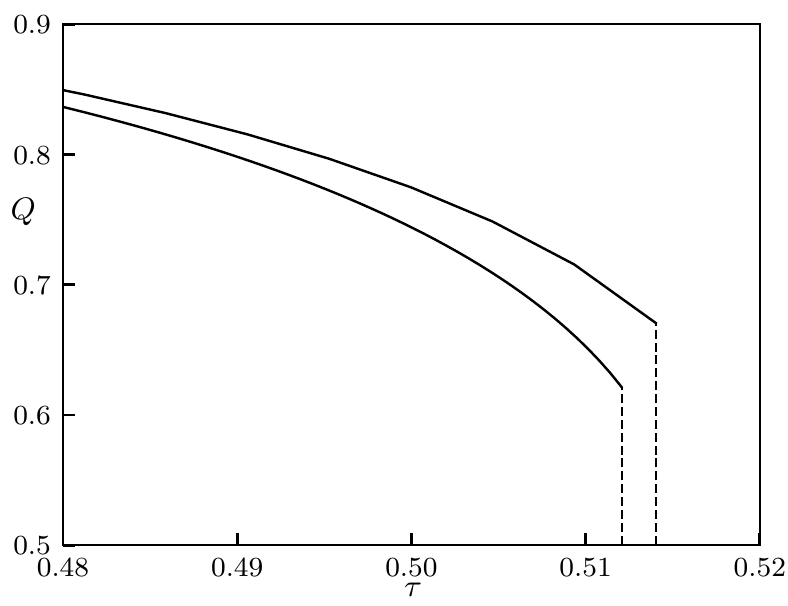}
    \caption{Nematic order parameter $Q$ as a function of 
    $\tau$ close
    to the coexistence temperature $\tau_c$. The upper curve
    corresponds to the triangle-Husimi lattice. The 
    lower curve is for the hexagon tree, calculating $Q$ 
    at the central site of the central hexagon, in the limit $\gamma/\mu \rightarrow 0$.}
    \label{fig:tauq}
\end{figure}

The nematic order parameter $Q$ is shown in Fig. \ref{fig:tauq} as a
function of the temperature for both treelike lattices. We note that
the transition temperature is reduced, and the jump in the
order parameter becomes smaller as we move from the triangle tree
to that with hexagons. Also, the coexistence temperature decreases and 
the temperature interval
in which both fixed points are stable becomes narrower. These results
are consistent, as the transition temperature on the triangular 
lattices estimated from the simulations is still lower and the transition is
continuous. 

\begin{figure}
    \centering
    \includegraphics[scale=1.]{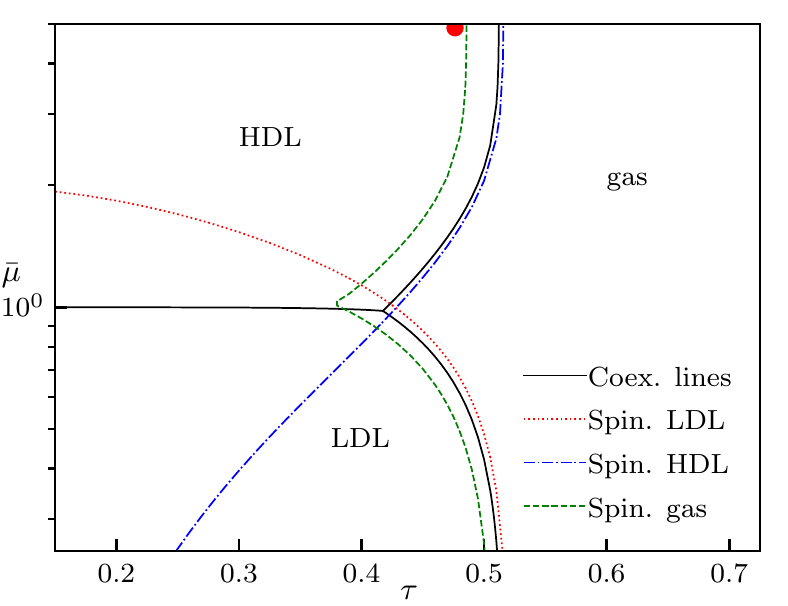}
    \caption{Phase diagram on the hexagon-Husimi tree, for finite chemical potential,
    in the ${\bar \mu} \times \tau$ plane. The three phases 
    (HDL, LDL, and gas) 
    are separated by coexistence lines (full lines in graph), 
    which meet at a triple point. Spinodal lines of the phases 
    are also shown. The red circle represents the critical 
    temperature obtained via MC simulations.}
    \label{fig:muxt}
\end{figure}

Finally, let us discuss how transition we study here fits into the general
phase diagram of the model for finite chemical potential. The phase diagram of the
complete model on the hexagon-Husimi tree is shown in Fig.
\ref{fig:muxt}, in the $\tau \times {\bar \mu}=\mu/\gamma$ plane. 
In this diagram we see that the gas-HDL coexistence line, as well as the gas and 
HDL spinodals, approach vertical asymptotes, and are very close to them already 
for $\bar{\mu} \gtrsim 10$, whose $\tau$ values are very close to the ones presented above for 
the solution in the full-occupancy case. Hence, our results for the infinite
chemical potential limit correspond to the final point of the coexistence line
between the HDL and gas phases. 
For finite $\mu$ the coexisting phases differ in the order parameter $Q$, and 
in the particle density $\rho$; the density is lower and $Q$ vanishes in the gas
phase. In the full-occupancy limit we consider here, $Q$ still vanishes in 
the gas phase and is finite in the HDL phase, but $\rho=1$ in both
phases.

\begin{figure}
    \centering
    \includegraphics[scale=1.]{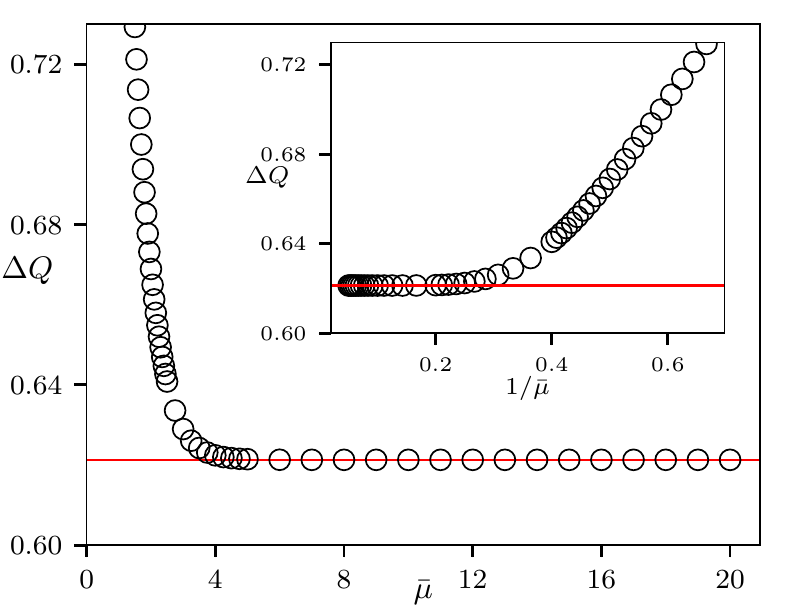}
    \caption{Discontinuity of the nematic order parameter at the 
    gas-HDL coexistence line as a function of the reduced 
    chemical potential ${\bar\mu}$. The insertion shows the same data against $1/\bar{\mu}$.}
    \label{fig:Dqxmu}
\end{figure}

Figure~\ref{fig:Dqxmu} shows the discontinuity of the nematic order 
parameter at the gas-HDL coexistence line as a function of the reduced 
chemical potential ${\bar \mu}$. We note that $\Delta Q$ becomes smaller 
as the chemical potential grows, and reaches its minimum value in the 
limit studied here. We can imagine three possible scenarios for 
this part of the phase diagram on the triangular lattice. 
One possibility is that in the limit $1/\mu \to 0$ the gas-HDL coexistence 
line ends at a critical point. Another would be that the transition 
becomes continuous already at some finite value of $\mu$, and thus the 
coexistence line would end at a tricritical point. Finally, the whole 
transition may be continuous. In this case, if the other transition lines 
(LDL-HDL and gas-LDL), which meet with the gas-HDL transition line at a 
triple point on the hexagon tree (see Fig. \ref{fig:muxt}), 
remain discontinuous, this point would become a critical end point.

\section{Conclusions}\label{sec:conc}

We investigate the gas-HDL phase transition of the ALG 
model\cite{He05} in its symmetric version\cite{Ba07} in the limit $1/\mu \to 0$ 
(full occupation) by Monte Carlo simulation and solution on Husimi lattices 
mean-field approximations for the solution on the triangular lattice).
The simulations reveal a continuous transition at 
$\tau_c=0.4763(1)$. 
The critical temperature is estimated using finite-size scaling estimates derived from the order parameters, their 
susceptibilities, the specific heat, and the fourth-order Binder cumulants. 
In addition, we derive an estimate of $U_4^*$ for the three-state Potts model on a square 
lattice: $U_4^*=0.6124(8)$. This value is close to, but 
more accurate than, one reported previously [$U_4^*=0.61$]~\cite{To02}. 
The critical exponents reported in Sec. IV also support a 3-state Potts model 
universality class. 

The mean-field calculations are performed on Husimi trees constructed with triangles and hexagons. 
Both approaches revel a discontinuous phase transition with coexistence temperatures somewhat higher than the simulation estimate, viz.,
$\tau_c=0.51403$ (triangles) and $\tau_c=0.51207$ (hexagons). In the second case the difference 
in relation to the simulation is $\lesssim 7\%$. We note that mean-field calculations for the three-state Potts model in two-dimensions also yield discontinuous transitions~\cite{Mi74}, which may explain the difference found here for the ALG model. 

The complete (mean-field) phase diagram shows a gas-HDL coexistence line which extends from $\bar{\mu} \approx 1$ to $\bar{\mu} \to
\infty$, along which a reduction of the discontinuity of the nematic order parameter $Q$ is observed as $\mu$ increases. The existence of the discontinuous line gives origin to three possible scenarios that connect the results of simulations and mean-field. The first is that at the limit $1/\mu \to \infty$ the line ends in a critical point, the second is a change in the order of transition at a finite chemical potential, and the third raises the possibility of the whole transition be continuous for the triangular lattice. We remark that the ALG model with finite $\bar{\mu}$ is somewhat similar to a diluted three-state Potts system, for which a continuous transition is also observed~\cite{Ni79,Qi05}. This suggests that the entire gas-HDL line might be continuous. Another indication of this is the fact that it is continuous in the original 2D ALG model~\cite{Sz09}, as well as in the 3D version~\cite{Bu08,Sz10}. 
In future work, we intend to perform Wang-Landau sampling, as well as transfer matrix calculations in order to obtain the complete phase diagram and identify which of the three scenarios mentioned above is correct.

\section{Acknowledgments}
The authors acknowledge financial support from the CNPq, CAPES and Fapemig, Brazilian agencies. 
We thank to M.C. Barbosa and W. Selke for helpful discussions and also Statistical physics 
Laboratory of Universidade Federal de Minas Gerais for the computational support.  

\appendix*

\section{Solution of the model on Husimi trees}
\label{app}

We show in some detail the calculations which lead to the 
thermodynamic properties on a Husimi tree, particularly
the one built with
triangles, as shown in Fig. \ref{fig:tree-tri}. As described above, we
start by defining partial partition functions (ppf's) for a sub-tree with 
a fixed configuration of the molecule on the root site. We thus define 
nine ppf's $g_{i,\eta}$ of rooted sub-trees, where the first index assumes
the values $1,2$ or $3$ of the direction of the edge opposite to the root
site of the triangle. On the sub-tree obtained removing the hatched
triangles in Fig. \ref{fig:tree-tri}, this index is 1. The
second index specifies the orientation $\eta$ of the non-bonding arms of
the molecule at the root.

The recursion relations are obtained by considering the operation of building a
sub-tree with an additional generation $(M+1)$ by connecting two pairs of
sub-trees (with $M$ generations each) to the vertices which are opposite to 
the root vertex of the new root triangle. We call $j_i$ the orientation of 
the molecule placed on the vertex opposite to the edge of orientation $i$ of 
the root triangle. The ppf's of a sub-tree with $M+1$ generations are thus
obtained through the recursion relations:
\begin{subequations}
\begin{eqnarray}
g^{(M+1)}_{1,j_1}&=&\sum_{j_2=1}^3\sum_{j_3=1}^3 W_{\{j_1,j_2,j_3\}}
g^{(M)}_{1,j_2}g^{(M)}_{1,j_3}
g^{(M)}_{2,j_3}g^{(M)}_{3,j_2}, \\
g^{(M+1)}_{2,j_2}&=&\sum_{j_3=1}^3\sum_{j_1=1}^3 W_{\{j_1,j_2,j_3\}}
g^{(M)}_{2,j_3}g^{(M)}_{2,j_1}
g^{(M)}_{3,j_1}g^{(M)}_{1,j_3}, \\
g^{(M+1)}_{3,j_3}&=&\sum_{j_1=1}^3\sum_{j_2=1}^3 W_{\{j_1,j_2,j_3\}}
g^{(M)}_{3,j_1}g^{(M)}_{3,j_2}
g^{(M)}_{1,j_2}g^{(M)}_{2,j_1}.
\end{eqnarray}
\label{rr}
\end{subequations}
The function $W_{\{j_1,j_2,j_3\}}$ is the statistical weight associated to
the edges of the root triangle, defined as
$W=\exp[2n_b(j_1,j_2,j_3)/\tau]$, where $n_b$ is the number of edges of 
the triangle 
with no non-bonding arms
on them and $\tau$ is the temperature. Thus:
\begin{eqnarray}
  n_b(&j_1&,j_2,j_3)=n_{b,1}(j_2,j_3)+n_{b,2}(j_3,j_1) \\ \nonumber
  &+& n_{b,3}(j_1,j_2) = \delta_{j_2,1}\delta_{j_3,1}+
  \delta_{j_3,2}\delta_{j_1,2}+\delta_{j_1,3}\delta_{j_2,3},
  \label{nb}
\end{eqnarray}
where $\delta_{i,j}$ is the Kronecker delta.  In this expression $n_{b,i}$
stands for the number of edges in direction $i$ with a bond on them, assuming
the values 0 or 1, so that, as expected, $n_b \in [0,3]$. Thus, the system of 
9 recursion relations [Eqs. (\ref{rr})] allows us to obtain the ppf's of
sub-trees with an arbitrary number of generations of triangles and, most
importantly, the thermodynamic limit (when $M \to \infty$). Usually, the ppf's
diverge in this limit, so it is convenient to define ratios of ppf's
\begin{equation}
  R_{i,j}=\frac{g_{i,j}}{g_{i,1}+g_{i,2}+g_{i,3}}.
  \label{ratios}
\end{equation}
These ratios generally converge to finite values in the thermodynamic limit; 
six of them are independent, since $\sum_j R_{i,j}=1$. Therefore, the
thermodynamic behavior of the model will be determined by the stable fixed
points of the recursion relations for the ratios $R_{i,j}$, which are obtained
from Eqs. (\ref{rr}).

We find fixed points with two different symmetries. In the isotropic fixed 
point the values of the 9 ratios, represented as a $3 \times 3$ matrix, are:
\begin{equation}
  {\mathbf R}_{iso}^*=
  \begin{pmatrix}
    x & (1-x)/2 & (1-x)/2 \\
    (1-x)/2 & x & (1-x)/2 \\
    (1-x)/2 & (1-x)/2 & x 
  \end{pmatrix},
\end{equation}
where the parameter $x$ is a function of the temperature $\tau$, whose exact
expression can be easily found with the help of an algebra software, but
is too long to be presented here. The three nematic fixed points are:
\begin{eqnarray}
  {\mathbf R}_{nem,1}^*&=&
  \begin{pmatrix}
    x_1 & (1-x_1)/2 & (1-x_1)/2 \\
    x_2 & x_3 & 1-x_2-x_3 \\
    x_2 & 1-x_2-x_3 & x_3 
  \end{pmatrix}, \\
  {\mathbf R}_{nem,2}^*&=&
  \begin{pmatrix}
    x_3 & x_2 & 1-x_2-x_3 \\
    (1-x_1)/2 & x_1 & (1-x_1)/2 \\
    1-x_2-x_3 & x_2 & x_3 \\ 
  \end{pmatrix}, \\
{\mathbf R}_{nem,3}^*&=&
\begin{pmatrix}
    x_3 & 1-x_2-x_3 & x_2 \\
    1-x_2-x_3 & x_3 & x_2 \\
    (1-x_1)/2 & (1-x_1)/2 & x_1 \\ 
  \end{pmatrix}.
\end{eqnarray}
Again the parameters $x_i$, with $i=1,2,3$, are functions of the temperature. We
note that the isotropic fixed point is stable at high enough temperatures, while
at lower temperatures one of the nematic fixed points is reached. 
We will see below that in the isotropic phase the orientation of the non-bonding
arms of the molecules has no preferred direction, while in the nematic phase one of
the three directions is more probable than the other two. As expected for a
discontinuous transition, the isotropic and nematic fixed points are both stable in 
an interval of temperatures around the coexistence temperature. 
The region of stability of a given fixed point may be found from the largest
eigenvalue of the Jacobian of the recursion relations for the ratios:
\begin{equation}
    J_{i,j}=\left(\frac{\partial R_i^\prime}{\partial R_j}\right)_{R^*},
\end{equation}
where $R_i^\prime$ and $R_i$ denotes the ratios in generations $M+1$ and $M$,
respectively, and the derivative is calculated at the fixed point whose stability 
is being considered. We find that the isotropic fixed point is stable for 
$\tau \geq 0.47299$, while the nematic ones for $\tau \leq 0.51970$. Hence, for 
$\tau \in [0.47299,0.51970]$ both fixed points are stable, so that the isotropic 
and nematic phases coexist in this temperature interval. This signals a
discontinuous transition between the phases and, then, the coexistence temperature
can be determined as the one at which the free energy in the bulk of the tree is
equal for both phases. 

Following the ansatz proposed by Gujrati\cite{Gu95}, we assume that the contribution to the
free energy per triangle is different for triangles located at the surface of the
tree ($\phi_s$) and the ones in the bulk ($\phi_b$). In a tree with $M$ generations
of triangles, the number of triangles in the bulk is $N_b=1+6+6\cdot 4+...+6 \cdot 4^{M-2}$,
while the number of triangles on the surface is $N_s=4\cdot4^{M-1}$. Therefore, if
$\varphi_M$ is the free energy of a tree with $M$ generations of triangles, we 
find that $\varphi_{M+1}-4\varphi_M=3\phi_b$, and since  $\varphi_M=-k_BT\ln Y_M$,
we have that:
\begin{equation}
  \phi_b=-\frac{1}{3}k_BT\ln \frac{Y_{M+1}}{Y_M^4},
  \label{phib}
\end{equation}
where $Y_{M}$ is the partition function on the whole tree, which can 
be obtained by connecting six sub-trees to the central triangle. This procedure
leads to:
\begin{equation}
  Y_{M}=\sum_{j_1,j_2,j_3=1}^3
  W_{\{j_1,j_2,j_3\}}g^{(M)}_{1,j_2}g^{(M)}_{1,j_3}g^{(M)}_{2,j_1}
  g^{(M)}_{2,j_3}g^{(M)}_{3,j_1}g^{(M)}_{3,j_2}.
  \label{y}
\end{equation}
In the thermodynamic limit $M \to \infty$, using the recursion relations [Eqs. (\ref{rr})], the ratios [Eqs. (\ref{ratios})] at the fixed points, and this
expression for the partition function [Eq. (\ref{y})], the bulk free energy per
triangle may be obtained through Eq. (\ref{phib}). This yields the coexistence
temperature $\tau_c=0.51403$. 
Therefore, we indeed have a {\em discontinuous} transition between the isotropic 
and nematic phases in the solution of the model on this Husimi lattice.

To further confirm this, we calculate the order parameter $Q$ as defined in Eq.
\ref{eq:defQ}. From the partition function (\ref{y}) it is simple to obtain the 
mean value of the number of molecules with orientation $i$ of non-bonding arms in
the central triangle:
\begin{eqnarray}
\langle n_i \rangle^{(M)}=\frac{1}{3
Y^{(M)}}\sum_{j_1,j_2,j_3=1}^{3} &n_i&
W_{\{j_1,j_2,j_3\}} \\ \nonumber
&\times& g^{(M)}_{1,j_2}
g^{(M)}_{1,j_3}g^{(M)}_{2,j_1}g^{(M)}_{2,j_3}g^{(M)}_{3,j_1}g^{(M)}_{3,j_2},
\end{eqnarray}
where this number has been normalized so that $0 \le \langle n_i \rangle \le 1$. 
If we divide the numerator and the denominator by
$\prod_{i=1}^3(g^{(M)}_{i,1}g^{(M)}_{i,2} g^{(M)}_{i,3})^2$ we may express this 
mean values in terms of the ratios, which in the thermodynamic limit assume their
fixed point values. The result is:
\begin{equation}
\langle n_i \rangle=\frac{\sum_{j_1,j_2,j_3=1}^3 n_i f(j_1,j_2,j_3,\{R^*\})}
{\sum_{j_1,j_2,j_3=1}^3 f(j_1,j_2,j_3,\{R^*\})},
\label{eq:f_i}
\end{equation}
where:
\begin{eqnarray}
f(j_1,j_2,j_3,\{R^*\}) &\equiv& W_{\{j_1,j_2,j_3\}} \\ \nonumber
&\times& R^*_{1,j_2} R^*_{1,j_3}R^*_{2,j_1}R^*_{2,j_3}R^*_{3,j_1}R^*_{3,j_2}.
\end{eqnarray}
As expected, in the isotropic gas phase, one has $\langle n_1 \rangle=\langle
n_2\rangle=\langle n_3\rangle$, so that the order parameter is $Q_{gas}=0$. On the
other hand, in the nematic HDL phase we find a non-vanishing $Q_{HDL}$, whose variation
with the temperature near the coexistence is displayed in Fig. \ref{fig:tauq}. At
the coexistence point the discontinuity in $Q$ is $\Delta Q = 0.67056$.

We performed similar calculations for the tree built with hexagons (shown in Fig.
\ref{fig:tree-hex}), which are actually a particular case of the ones presented 
in \cite{Ol11}, when all lattice sites are occupied by molecules. For
brevity, we will not detail the calculations here, and present the main results
only. Since the tree in this case is built with larger clusters, hexagons with one
central site and six sites at the border, it is expected that such a calculation
should lead to results which are closer to those on the triangular lattice.
Again, we find a discontinuous transition, at a somewhat lower temperature,
$\tau_c=0.51207$. The discontinuity in the nematic order parameter
at the coexistence is $\Delta Q=0.6216$, and for $\tau \in [0.48569,0.51575]$ both
fixed points are stable. In this case, since there are effectively four sites per
hexagon on the treelike lattice ($6/2$ at the hexagon's border and the central one),
there are more than one way to define the order parameter at the central hexagon.
One could, for example, consider just the central \textit{site} or a mean value of
the order parameter over all sites of the central \textit{hexagon}. We have done
both calculations and found out that these values are quite close. The deviation 
is maximum at the coexistence temperature and the relative difference is about
0.25\% there.

\bibliography{references.bib}

\end{document}